# Impact of free electron degeneracy on collisional rates in plasmas

Gareth O. Williams,[1,*] H.-K. Chung,[2,3] S. Künzel,[1] V. Hilbert,[4] U. Zastrau,[5] H. Scott,[6] S. Daboussi,[7] B. Iwan,[8] A. I. Gonzalez,[8] W. Boutu,[8] H. J. Lee,[9] B. Nagler,[9] E. Granados,[9] E. Galtier,[9] P. Heimann,[9] B. Barbrel,[10] R. W. Lee,[6] B. I. Cho,[2] P. Renaudin,[11] H. Merdji,[8] Ph. Zeitoun,[1] and M. Fajardo[1]

[1]*GoLP/Instituto de Plasmas e Fusão Nuclear-Laboratório Associado, Instituto Superior Técnico, Universidade de Lisboa, 1049-001 Lisboa, Portugal*
[2]*Department of Physics and Photon Science, Gwangju Institute of Science and Technology, Gwangju 61005, Korea*
[3]*National Fusion Research Institute, Daejeon 34133, Republic of Korea*
[4]*Institute of Applied Physics, Friedrich Schiller University Jena, Albert-Einstein-Strasse 6, 07745 Jena, Germany*
[5]*European XFEL, Holzkoppel 4, 22869 Schenefeld, Germany*
[6]*Lawrence Livermore National Laboratory, Livermore, California 94550, USA*
[7]*Laboratoire d'Optique Appliquée, ENSTA, CNRS, IP Paris, 181 Chemin de la Hunière et des Joncherettes, 91762 Palaiseau cedex, France*
[8]*LIDYL, CEA, CNRS, and Université Paris-Saclay, CEA Saclay, 91191 Gif-sur-Yvette, France*
[9]*SLAC National Accelerator Laboratory, 2575 Sand Hill Road, Menlo Park, California 94025, USA*
[10]*Lawrence Berkeley National Laboratory, 1 Cyclotron Road, Berkeley, California 94720, USA*
[11]*CEA-DAM-DIF, Bruyères Le Châtel, 91297 Arpajon Cedex, France*

Degenerate plasmas, in which quantum effects dictate the behavior of free electrons, are ubiquitous on earth and throughout space. Transitions between bound and free electron states determine basic plasma properties, yet the effects of degeneracy on these transitions have only been theorized. Here, we use an x-ray free electron laser to create and characterize a degenerate plasma. We observe a core electron fluorescence spectrum that cannot be reproduced by models that ignore free electron degeneracy. We show that degeneracy acts to restrict the available electron energy states, thereby slowing the rate of transitions to and from the continuum. We couple degeneracy and bound electron dynamics in an existing collisional-radiative code, which agrees well with observations. The impact of the shape of the cross section, and hence the magnitude of the correction due to degeneracy, is also discussed. This study shows that degeneracy in plasmas can significantly influence measured observables such as the emission spectra, and that these effects can be included parametrically in well-established atomic physics codes. This work narrows the gap in understanding between the condensed-matter and plasma phases, which coexist in myriad scenarios.

## I. INTRODUCTION

Many properties of plasmas, such as the degree of ionization, stopping powers, and the transport of energy, are shaped by electron transitions involving the free electron continuum. These transitions define basic plasma properties such as the degree of ionization. The light emitted during electron transitions in plasmas (the emission spectrum) provides a fingerprint of the plasma conditions and is arguably the most widely used plasma diagnostic. The rates of these transitions determine the emission spectrum, and accurate knowledge of these rates is of paramount importance to understand experimental observations and predict the behavior of high-energy-density (HED) matter found throughout the universe and in HED laboratories worldwide. In classical plasmas, the number of free electron states in the continuum is unbounded and thermal free electrons are well described by Maxwell-Boltzmann (MB) statistics. This description allows for a straightforward inclusion of free electrons in a multitude of theoretical treatments, including ionization and recombination. However, this assumption is not valid in plasmas where the free electrons are governed by quantum effects (degenerate). Quantum mechanics limits the available electron states at a given energy level through the exclusion principle, resulting in the free electrons obeying Fermi-Dirac (FD) statistics. A consequence of the exclusion principle is that fully occupied states cannot accept additional electrons, and this transition is blocked (Pauli blocking) [1]. Therefore, degeneracy influences electron transitions to or from the continuum by both shaping the free electron energy distribution and by blocking transitions to occupied states. Electron degeneracy is present in numerous HED scenarios, such as collapsing stars [2,3], the compression phase of inertial confinement fusion (ICF) plasmas [4], shock compressed plasmas [5], the centers of large planets [6], and the early stages of all high-intensity optical or x-ray laser interactions with solids [7,8]. Furthermore, degenerate plasmas can exist at any temperature, provided the electron density is sufficiently high.

The statistics of a degenerate free electron distribution was first treated by Fermi and Dirac in 1926 [9,10]. Soon after, these quantum statistics were applied to explain several previously anomalous properties of metals in an approach called the *free electron model* [11]. This approach hinges on the assumption of a noninteracting electron gas embedded in a structureless background of positive charge, and can account for thermal and electrical conductivities and the density of states (DOS) for simple metals. However, this treatment does not explicitly include bound electrons, which are necessary for modeling emission spectra.

A model that can include a quantum treatment of all electrons for any element at a given temperature and density is the *average atom* (AA) model [12]. This approach calculates the average wave functions, average energy levels, and average occupations of a single atom surrounded by free electrons. This approach has proven extremely successful at calculating macroscopic properties such as tables for a range of elements [13,14]. Furthermore, the average wave functions can be used to calculate various ionization and recombination cross sections. The drawback to this method is that the spectrum calculated from the average atom does not exhibit the detailed spectra we observe in nature. For example, the various lines of bound-bound electron transitions in a plasma are due to photon emission from atoms with discrete electronic configurations, with each atom emitting a unique line. The ratio of these lines contains valuable information about the charge states and electron temperatures of the plasma. In the average atom framework, only a single average charge state is represented that emits an average spectral line for a given bound-bound transition. This makes the approach of limited utility for modeling detailed emission spectra [15]. Some hybrid approaches have shown that more accurate levels and wave functions corresponding to specific unitary (non-partial) electron configurations are possible, which may aid in the development of more accurate cross sections [16].

Arguably the most successful treatment of electron quantum effects such as degeneracy of arbitrary atomic elements and structures is density functional theory (DFT) [17]. DFT calculates the most likely electron density of an arrangement of atoms and represents the density with a set of single-electron wave functions (Kohn-Sham orbitals). The main advantage of DFT over the AA technique is the ability to include structural effects, which take the form of ion-ion correlations in a plasma. The electron interactions are approximated with the exchange-correlation (xc) functional in DFT. Bound electrons can be included in DFT, which can yield the complete electronic structure of complex systems in their ground state. DFT was extended to finite temperatures by Mermin, making it applicable to HED conditions [18]. Much like the AA approach, thermal DFT uses partial occupancies to account for the temperature, therefore only yielding average wave functions and occupations. To estimate more exact orbital energies in DFT requires the inclusion of predefined ions in a given plasma environment [19]. Even in the absence of exact treatments for the temperature-dependent exchange-correlation (xc) functional in DFT [20,21], it has proven remarkably successful at modeling a wide range of HED plasma properties such as ion structure [5], ionization potential depression (IPD) [19], the optical properties [22], conductivity [23], and the equation of state (EOS) [24]. Despite the capability to capture many properties of HED plasmas, DFT *a priori* cannot capture time-dependent processes such as collisional ionization and recombination that define the emission spectrum observed in experiments.

An extension to DFT that includes time-dependent effects is time-dependent DFT (TDDFT) [25]. It can be applied to model laser excitation of electrons in solids [26], photoionization rates [27], stopping powers of ions [28], or the x-ray scattering properties of HED plasmas [29]. However, to include the rapid transfer of energy between charges, reliable nonadiabatic xc functionals are required, to which approximations are only available for limited cases [30]. The approximation of electron-electron interactions present in DFT also applies to TDDFT, and therefore excludes the electron-electron processes needed to calculate electron transitions [31]. Attempts to model binary collisions (one electron colliding with a H atom) with TDDFT have shown that it is unable to replicate the results of exact treatments, showing that TDDFT in its current form is unable to account for collisional ionization or recombination [32,33]. To model the detailed emission spectrum stemming from the multitude of energy levels and electron transitions present in realistic plasmas, a model that can calculate all of these transitions in a wide range of plasma conditions is necessary.

Time-dependent collisional-radiative (CR) codes, initially designed for hot, nondegenerate plasmas, are the tools of choice for simulating emission spectra [34]. Given a certain temperature and ion density, CR codes use tabulated or measured electron levels, ionization cross sections and oscillator strengths, from which the rates of all transitions are calculated. These rates determine the population of the levels, which can be calculated in a time-dependent manner. The energy levels and cross sections can, in principle, be included up to an arbitrary level of complexity. The emission spectrum can then be calculated using the rates of photon-emitting processes and plasma opacity effects. These codes can be coupled to hydrodynamic simulations that supply the temperature and density to the CR code, which then calculates a spectrum that is emitted from a highly dynamic and nonequilibrium system, such as laser-driven ICF [15]. However, the atomic data available for CR codes do not include interactions of neighbouring charges on the free or bound electrons. It is precisely these interactions that are responsible for degeneracy, continuum lowering, and line broadening effects. Since these high density effects are not present in the available data, models or parameterized descriptions of them must be added as part of the CR approach. This approach has already been used to account for continuum lowering or IPD in solid-density plasmas [19,35,36], although which model is more accurate is debated [37,38]. Degeneracy effects such as Pauli blocking can influence ionization potential depression, further highlighting the need to include these effects in CR codes [39]. A tractable method for including free electron quantum effects such as degeneracy in CR codes is pivotal in bridging the gap between the often coexisting condensed-matter and plasma phases.

In this article, we present x-ray fluorescence spectroscopic measurements of a degenerate solid-density aluminium plasma to elucidate free electron degeneracy effects on atomic transitions. The measured fluorescence spectra show marked differences to those predicted by models using a classical MB treatment of the free electrons. We show that by correcting the rates of atomic processes in a CR code using FD statistics to account for degeneracy, a good match with the data can be found. The effect of the cross section used in calculating the degeneracy corrections is also investigated. Calculated spectra show a better agreement with experiment when a cross section favoring lower electron energies is used. The effects of degeneracy accentuate the effects of different cross-section profiles, making them experimentally distinguishable.

## II. EXPERIMENTAL APPROACH

Theoretical treatments predict a substantial (several orders of magnitude) reduction of the rates of collisional ionization and recombination with increasing degeneracy, yet experimental confirmation is so far lacking [34,40,41]. To test these predictions, the rate of collisional ionization or recombination must be measured in a plasma of known degeneracy. Experimentally, this presents a significant challenge: first, collisional ionization of solids requires temperatures well above those in the degenerate regime at solid density, making a transient nonequilibrium measurement necessary, and second, the level of degeneracy (electron temperature and density) must be well defined during the measurement.

The intensities and pulse durations now available at x-ray free electron laser (XFEL) facilities have naturally led to the creation of plasma states with well-defined temperatures and densities from solid targets [42,43]. Uniquely, the ultrashort nature of XFEL pulses (typically <100 fs) and tuneable photon energy allow the x-ray fluorescence signature of the plasma to be measured before significant changes in the ion density occur (due to target expansion) [44]. Similar techniques have facilitated the measurement of ionization potential depression [19,35,36], collisional rates [45,46], saturable [47,48] and reverse-saturable [49] x-ray absorption, and x-ray opacities of solid-density plasmas [50]. The XFEL pulse can be tailored to create a small population of high-energy electrons to drive ionization in an otherwise degenerate plasma background, while simultaneously providing the XFEL-induced fluorescence spectrum characteristic of the level of ionization, making it an ideal tool for investigating degenerate plasmas.

We use the Linac Coherent Light Source (LCLS) XFEL [51] and the Matter in Extreme Conditions instrument [52] to deliver pulses of 37 $\mu$J in energy, 35 fs in duration at full width at half maximum, with a photon energy of 3 keV, to heat solid foils of aluminium of 300 and 600 nm in thickness. The XFEL beam is focused with a stack of beryllium lenses to a spot of about 25 $\mu$m in diameter. The pathway of XFEL-induced fluorescent emission starts with the absorption of an incoming photon of $E^{XF} = 3$ keV, exciting a $K$-shell electron at a binding energy of $E^K \approx 1560$ eV to an energy of $E^{XF} - E^K \approx 1440$ eV (where $K$, $L$, and $M$ correspond to the first, second, and third electron shells, respectively). The subsequent filling of the $K$-shell hole occurs via an $L$-shell electron relaxing to the vacant $K$-shell hole. In aluminium, this $L$ to $K$ shell transition releases an energy of $E^{LK} \approx 1485$ eV and results in the emission of an Auger electron at an energy of $E^A \approx E^{LK} - E^L \approx 1400$ eV, where $E^L$ is the binding energy of the $L$-shell electron. Auger emission accounts for about 96% of the recombination events, leaving two holes in the $L$ shell. The other 4% of recombination events relax via the emission of a photon, and these photons constitute the measured fluorescence spectrum.

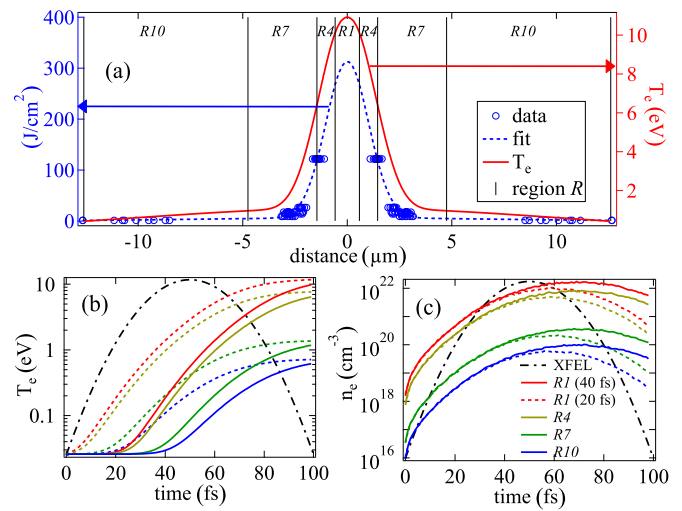

FIG. 1. (a) The fluence scan data (blue circles) with fitted curve (dashed blue line) and corresponding peak electron temperatures (solid red line) of the XFEL focal spot radial profile. The spot is separated into 10 radial volumes for modeling, labeled $R$, that absorb the same number of XFEL photons, and four such regions are shown by vertical black lines. (b) The XFEL pulse intensity profile in time (black dash-dotted curve) and corresponding electron temperatures for two heating delays of 20 fs (dashed curves) and 40 fs (solid curves) for four plasma volumes from $R1$ (top curves) to $R10$ (bottom curves). (c) The corresponding hot electron densities, $n_e$.

The fluorescence of the photoionized solid-density plasma was recorded using an x-ray spectrometer with a 001-orientation rubidium hydrogen phthalate (RHP) crystal of 100 $\mu$m thickness placed on a curved substrate with a 5 cm radius of curvature that focused the fluorescent emission onto an x-ray detector. An aluminium filter of 1.6 $\mu$m thickness was used at the spectrometer entrance to reject optical light. The spectral resolution of the system was ~1 eV. More than 100 shots were accumulated for each measurement. To infer the plasma conditions, the XFEL spot profile was measured using imprints of the ablation pattern for a range of fluence values (F-scan), as shown in Fig. 1(a) [53,54]. This measurement, when combined with an XUV emission spectrum measurement, shown in Ref. [22], allowed us to confirm a peak electron temperature of ~12 eV in the hottest, central region of the plasma [$R1$ in Fig. 1(a)]. Since the total pulse energy is spread over a much larger region, we divide the plasma into 10 concentric volumes, each of which absorbs the same number of XFEL photons [we select regions *R*1, *R*4, *R*7, and *R*10 for clarity, shown in Fig. 1(a)].

## III. PLASMA CONDITIONS AND CORRECTIONS TO THE RATES

The absorption of XFEL photons results in a population of high-energy photoionized and Auger electrons with energies of ~1.4 keV, that will be referred to as hot electrons. The electrons that constitute the cold-solid free electrons remain thermalized and will be referred to as the bulk electrons. To estimate the degeneracy of the bulk electrons, as characterized by $\Theta = \mu/(k_B T_e)$, where $\mu$ is the chemical potential of the plasma and $T_e$ the bulk electron temperature, we must model the temperature as a function of time. The bulk electrons (and hence the degeneracy) are initially unaffected by the XFEL photons and heat up only after a time delay relative to the XFEL pulse. We define this delay as the difference in time between the XFEL pulse and bulk electron heating, meaning a time delay of zero would correspond to instant thermalization of the XFEL energy into the bulk electrons. We have used delay times in our calculations which approximately agree with recently reported thermalization times of XFEL-generated hot electrons in silicon [55].

The time evolution of the bulk electron temperature within a selection of the volumes labeled *R* are shown in Fig. 1(b) for two separate time delays, i.e., 20 and 40 fs. Details of the bulk electron temperature calculations are reported elsewhere [22,56]. The corresponding hot electron density is shown as a function of time in Fig. 1(c). Modeling of the bulk electron temperature shows that the plasma remains degenerate throughout the XFEL pulse (as $T_e < E_F$, where $E_F \approx 12$ eV is the Fermi energy of aluminium at room temperature). The XFEL radiation, bulk electron temperatures, and hot electron densities in Figs. 1(b) and 1(c) are used as input into a CR code to model the fluorescence spectrum, as discussed later.

The bulk electrons never reach a temperature that is sufficient to cause appreciable ionization of the *L* shell. On the other hand, the hot electrons can repeatedly collisionally ionize the *L* shell until their energy is below that of the ionization threshold. Ultimately, the energy of the hot electrons is transferred to the bulk electrons, leading to increased $T_e$ [see Figs. 1(b) and 1(c)]. Ionization of the *L* shell by the hot electrons causes a reduction in screening and increased *K*-shell binding energy, resulting in $K\alpha$ fluorescence at higher photon energies. We will refer to these spectral lines as $K\alpha$ satellites and they are directly related to the number of vacancies in the *L* shell at the time of photon emission, irrespective of the exact electron configuration in the *L* shell. For the satellites to be observed in the measured spectrum, holes in the *L* shell must be present at the time the fluorescent *L* to *K* transition takes place. In this way, the fluorescent emission spectrum reflects the *L*-shell vacancies present only within the temporal envelope of the XFEL pulse.

To quantify the degeneracy effect on atomic rates, a correction factor (*CF*) is defined as the ratio of the atomic rate including degeneracy, $A^D$, to that without degeneracy, $A^C$, $(CF = \frac{A^D}{A^C})$. The *CF* values are calculated following theoretical frameworks presented elsewhere and are plotted as a function of degeneracy, $\Theta$, in Fig. 2 (see caption for details) [34,40,41]. The two processes dominating the number of *L*-shell vacancies in time are collisional ionization (CI) and its inverse process, three-body recombination (3BR), which are shown as dashed and solid curves in Fig. 2, respectively. The range of *CF* values corresponding to the plasma conditions at the peak of fluorescent emission for charge states V and VI (~60 fs) for heating delays of 20 and 40 fs are indicated in Fig. 2 by dash-dotted lines with arrows.

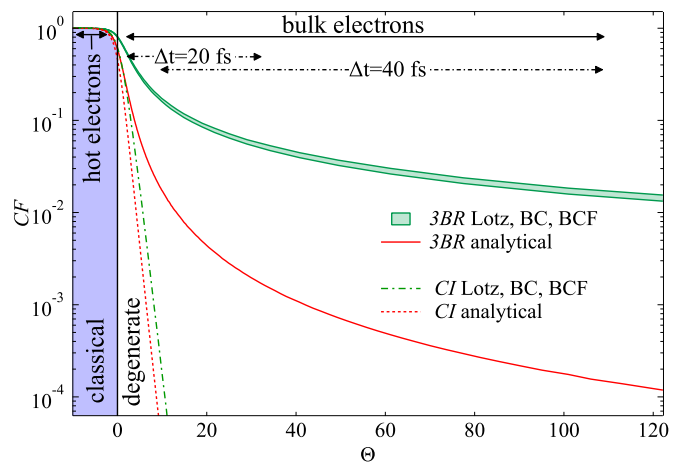

FIG. 2. The correction factor (*CF*) plotted against the plasma degeneracy $\Theta = \mu/(k_B T_e)$ at solid density. *CF* is shown for three-body recombination (green shaded curve) and collisional ionization (dash-dotted green curve) for three cross sections for a transition of energy $dE \gg T_e$ (details in text). The *CF* values for the analytical cross sections are shown for three-body recombination (solid red curve) and ionization (dashed red curve). The range of *CF* values for the hot and bulk electrons is shown by solid black arrows. The hot electrons are unaffected by degeneracy, and $CF \approx 1$. Conversely, the bulk electrons remain degenerate throughout the XFEL pulse. The *CF* values for the bulk electrons corresponding to the two heating delays used in Fig. 1 are indicated by black dash-dotted lines with arrows.

Calculating the rate, *A*, of a collisional process requires integrating the product of the electron distribution, $f_e(E)$, and the cross section, $\sigma(E)$, for that process, over electron energy, $E$ ($A \propto \int \sigma(E) f_e(E) dE$). As the rates depend on the choice of the cross section, $\sigma(E)$, we explore the effect of several cross-section models on the values of *CF* for comparison. The correction factor can then be applied straightforwardly to the classical MB rate as $CF \times A^C$, for a given level of degeneracy, $\Theta$.

First we calculate *CF* for several collisional-ionization cross sections commonly used in CR codes, namely, Burgess and Chidichimo (BC) [57], the modified Burgess and Chidichimo (BCF) [46], and Lotz [58], which are shown as the shaded green curve for 3BR and the dashed green curve for CI in Fig. 2. These expressions give values for total cross sections as a function of the energy, *E*, of the ionizing electron. Calculating *CF* for a degenerate electron distribution requires knowing the differential cross section as a function of both (outgoing) electron energies, as separate blocking factors apply to these electrons. However, models or experiments of such differential cross sections for degenerate plasmas are absent. The green curves in Fig. 2 assume the cross section of the outgoing electron is constant, i.e., the cross section is independent of the energy of the outgoing electrons.

As three cross sections shown by the green curves in Fig. 2 have similar shapes and yield similar values of $CF$, we include a simplified total cross section, with a $1/E$ dependence, to explore the influence of the cross-section shape on the correction factor $CF$. This $1/E$ dependence also permits the values of $CF$ to be calculated in a much simplified, semi-analytical manner, as detailed in Ref. [40]. We refer to this as the analytical cross section and apply the values of $CF$ calculated using this approximation to the rates of SCFLY using the other cross sections. The corresponding values of $CF$ are shown in red (analytical) in Fig. 2. Use of the analytical cross section results in values of $CF$ that are roughly an order of magnitude lower than the BC, BCF, and Lotz values of $CF$. This is due to the strong weighting of the analytical cross section $(1/E)$ on lower electron energies, where the differences between the FD and MB distributions at low temperatures are maximized and where blocking factors become important.

A collisional-ionization rate is calculated in SCFLY according to the chosen cross section, and then the 3BR rate is calculated using the detailed balance relationship. These rates are then corrected by the corresponding correction factors. As such, the balance between the two processes that govern the occupation of the $L$ shell (collisional ionization and recombination) determines the level of ionization at any given time. The cross section used to calculate the ionization rate does play a role, as is evident from the range between the shaded regions in Fig. 2. However, this range is small when compared to the corrections from degeneracy.

In the present study, the collisional ionization is driven by the hot electrons that have energies orders of magnitude greater than the Fermi energy. After the collision, the ionized and scattered electrons are therefore born at energies for which Pauli blocking is negligible. Conversely, three-body recombination occurs through the degenerate bulk electrons relaxing to vacant bound states. Therefore, the degeneracy correction to the collisional ionization by hot electrons is negligible $(CF \approx 1)$ and we can isolate the dominant process affected by degeneracy as three-body recombination. Contrary to collisional-ionization cross sections, models or measurements of three-body recombination cross sections are greatly lacking and its shape is not accurately known. Consequently, the three-body recombination cross section must be approximated from the collisional-ionization cross section through the microreversibility condition [59].

## IV. RESULTS AND DISCUSSION

Two XFEL-induced $K$-shell fluorescent spectra for target thicknesses of 300 and 600 nm are shown in Fig. 3 (solid back noisy lines). A negligible difference between the normalized spectra of both thicknesses is found, showing that the plasma is optically thin for the range of photon energies shown in Fig. 3. The three peaks IV, V, and VI visible in Fig. 3 correspond to the fluorescent $(L$ to $K)$ transitions with one, two, and three $L$-shell vacancies, respectively. Qualitatively, the appearance of satellites in the spectra signify that there are $L$-shell holes present at the time of fluorescence, which occurs during the XFEL pulse.

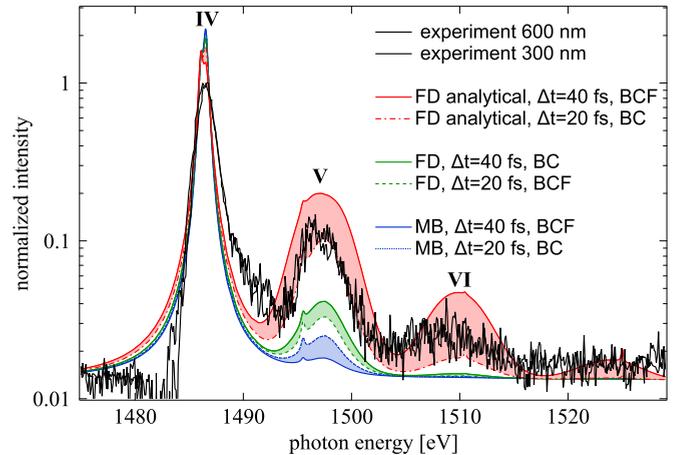

FIG. 3. The experimental $K\alpha$ fluorescence spectrum of XFEL irradiated aluminium showing the three ionization stages IV, V, and VI for 300 and 600 nm foil thickness (noisy solid black lines). Each shaded region is calculated using a different treatment of $CF$ according to Fig. 2. The spectra of two heating delays, and three cross sections are calculated for each treatment of $CF$. The spectra using MB statistics and no correction (shaded blue region with short-dashed line: bottom), FD statistics with the standard $CF$ values (shaded green region with dashed line: middle), and FD statistics with the analytical approximation (shaded red region, with dash-dotted line: top) are shown.

To quantitatively compare our measured spectrum with theoretical predictions, we use the time-dependent nonlocal-thermodynamic-equilibrium (non-LTE) CR code SCFLY [60–62], which includes the effects of both degenerate and nonthermal electron distributions. SCFLY has shown to be a robust tool for the analysis of similar experiments, albeit at higher temperatures [19,35,44–46,48–50]. Evolving a non-thermal free electron distribution that is fully self-consistent with all of the atomic processes is beyond the scope of this article. Instead, we approximate the temporal evolution of the hot electron distribution to be in broad agreement with other studies [63,64]. As the change in electron density is minimal, the spectra are insensitive to the ionization potential depression model that is used [19,35,36]. The bulk electron temperature [Fig. 1(b)], hot electron density [Fig. 1(c)], and hot electron distribution are input into SCFLY for every time step.

Spectra calculated using SCFLY with MB and FD statistics and different cross-section models are shown in Fig. 3 (shaded curves). The spectra using classical MB statistics in Fig. 3 (blue shaded curve) show a cold IV peak and a negligible V peak that do not agree well with the experimentally observed spectra. A standard interpretation of the ratio of the satellite intensities suggests a plasma with an average electron temperature of $T_e \approx 20$ eV. This disagrees sharply with the average $T_e$ of a few eV's from energy deposition measurements in Fig. 1, and a peak of $T_e \approx 13$ eV obtained from soft x-ray emission spectra [22]. Having tested a range of time delays and cross sections, we could not obtain good agreement with the data when using MB statistics. As the spectra of hot aluminium plasmas have been shown elsewhere [44] to be well reproduced by SCFLY in the classical regime, the discrepancy between the experimental data and spectra calculated assuming classical statistics here shows that reconsidering the rates in the degenerate regime is essential.

Adding the correction factors for the common cross sections (BC, BCF, and Lotz) to the collisional rates in SCFLY produces more prominent satellites and a much improved fit to the data (see Fig. 3, shaded green curve). Although a good match to the data is still lacking with these corrections, the main effect of decreased three-body recombination rates is evident from the increased prominence of the $K\alpha$ satellites. As the three-body recombination rate is suppressed, the $L$-shell holes persist longer than they would in the absence of degeneracy, causing more prominent satellite peaks. Since these cross sections have performed well in classical plasmas, we then apply the values of $CF$ calculated using the analytical approximation to these rates in SCFLY. This yields a good fit to the data, in which the difference between spectra calculated using the cross sections and delay times lies within the error of experimental data. Although the $1/E$ shape of the analytical cross section is an approximation, these results strongly suggest that correction factors for three-body recombination are better approximated with cross sections favoring lower electron energies.

## V. CONCLUSIONS

We have used the simultaneous pump-and-probe method [44] to create and diagnose a degenerate solid-density aluminium plasma with an XFEL and have shown that the resulting fluorescence spectrum is not reproduced with standard treatments that assume a classical MB free electron distribution. We identified the driving factor that causes differences between the calculated MB and FD spectra to be the lowered three-body recombination rate. By correcting the collisional rates in SCFLY to account for degeneracy, we demonstrated an improved fit to the data when using standard cross sections, yet a good fit was still absent. Using a simple $1/E$ cross section weighted towards lower electron energies gives larger reductions to the rates and the resulting spectra show a good fit to the data. This highlights that the precise shape of the 3BR cross section, especially at lower electron energies, may be poorly approximated in this degenerate regime. This work has shown experimentally that degenerate plasmas exhibit collisional rates that differ significantly from classical predictions. The method of using correction factors to the rates is applicable to any collisional-radiative code, and future refinements of cross sections and/or correction factors will allow more accurate calculations of degenerate plasma properties and resulting emission spectra. The results here will lead to a better understanding and greater control of a wide variety of degenerate and nonequilibrium HED states.

## ACKNOWLEDGMENTS

Use of the Linac Coherent Light Source (LCLS), SLAC National Accelerator Laboratory, is supported by the US Department of Energy, Office of Science, Office of Basic Energy Sciences under Contract No. DE-AC02-76SF00515. The MEC instrument is supported by the US Department of Energy, Office of Science, Office of Fusion Energy Sciences under Contract No. SF00515. We acknowledge support from Fundação para a Ciência e a Tecnologia (FCT) Projects No. EXPL/FIS-OPT/0889/2012, No. PTDC/FIS/112392/2009, and No. 02/SAICT/2017/31868, COST Action CA17126, Laserlab-Europe EU-H2020 Grant No. 654148, and the Swedish Foundation for International Cooperation in Research and Higher Education (STINT). This work was supported by the European Union's Horizon 2020 research and innovation programme (VOXEL H2020-FETOPEN-2014-2015-RIA 665207). H.-K.C. and B.I.C. acknowledge support for the National Research Foundation (Grants No. NRF-2015R1A5A1009962 and No. NRF-2016R1A2B4009631) of Korea. The work of H. Scott was performed under the auspices of the US Department of Energy by Lawrence Livermore National Laboratory under Contract No. DE-AC52- 07NA27344.